\documentclass{appolb}
\usepackage{graphicx}

\begin{document}
\title{
Measurement of the Transition Form Factor in \\ 
$\phi \to \pi^0 e^+ e^-$ with the KLOE detector
%
}
\author{Matteo Mascolo
\address{on behalf of the KLOE-2 collaboration}
\\
}
\maketitle

\begin{abstract}
By studying the invariant-mass distribution of the $e^+e^-$ in conversion decays, it is possible to learn more about mesons structure and underlying quark dynamics. At KLOE, the study of the $\phi \to \pi^0 e^+e^-$ process is currently going to be finalized. At present, about 9000 events have been selected from a data sample of 1.7 fb$^{-1}$ from 2004/2005 data taking campaign. A very good agreement between data and Monte Carlo distributions has been achieved for all interesting kinematical variables. A preliminary invariant-mass spectrum of $e^{+}e^{-}$ is presented.
\end{abstract}

  
\section{Introduction}

Transition Form Factors (TFF) are fundamental quantities needed for a detailed understanding of the nature of mesons and their underlying quark and gluon structure. They play an important role in many fields of particle physics, such as the calculation of the hadronic Light-by-Light contribution to the Standard Model value of the anomalous magnetic moment of the muon and the search for quark-gluon plasma in heavy ion collisions. Moreover, meson TFF represent a strong ``benchmark'' for theoretical modelling of different processes, being a field in which high precision measurements are possible.

In particular, the most important theoretical advances in modelling the conversion decay of a light vector resonance (V) into a light pseudoscalar meson (P) and a lepton pair $l^+l^-$
\begin{equation}
\label{eq:conv}
V \to P \,\, \gamma^{*} \to P \,\, l^+l^-
\end{equation}
were mostly driven by the $\sim\!\!10\,\sigma$ discrepancy between the experimental data of NA60 \cite{na601} and Lepton G \cite{lepg}, and the Vector Meson Dominance (VMD) ansatz \cite{VMD} prediction for the $\omega \to \pi^0 \gamma^{*}$ transition for factor.

In recent years, several theoretical models have been developed to justify this discrepancy \cite{mod1, ivs, mod3}. In this scenario, a measurement of the pion TFF in Dalitz decays of $\phi$ ($\phi \to \pi^0 \gamma^* \to \pi^0 e^+e^-$), which was never measured so far, is extremely valuable, since it would allow to expand the range of explored $q^2$ (the squared 4-momentum of the virtual photon) to the $\rho$ resonance mass region. 

The interest of the KLOE \cite{kloe} collaboration in $\phi \to \pi^0 e^+e^-$ is hence justified by the possibility to measure $F_{\pi^0}(m_{\phi}^2, q^2)$ in a kinematical region which fully includes the resonance enhancement, as well as the improvement the branching ratio measurement, currently known with an accuracy of $25\%$ in the PDG world average \cite{pdg}. Additional theoretical issues related to the study of this process, are described in \cite{ivs}.

\section{The analysis}

The analysis of the $\phi \to \pi^0 e^+e^-$ is performed on a 1.7 fb$^{-1}$ data sample collected at DA$\Phi$NE collider ($\sqrt{s} = m_{\phi}$), during the 2004-2005 KLOE data taking campaign. The signal was simulated according to the Landsberg $e^+e^-$ mass-spectrum distribution \cite{lands}, including different TFF parametrizations. The Monte Carlo procedure accounts also for both the initial and the final state radiation photon emissions. 

Due to a similar signature, signal events are mostly tagged as radiative Bhabha interactions by the event classification algorithm of KLOE. This results in a considerable background contamination from Bhabha events at the pre-analysis level\footnote{The pre-analysis level includes trigger and event classification selections, a machine-background filter (FILFO) and the requirement of exactly two tracks of opposite charge and two prompt neutral clusters coming from the interaction point.}. A huge fraction of this background can be eliminated requiring the following set of cuts: 

\begin{itemize}
\item{the lepton energies in the range: 30 MeV $< E_{e^+,e^-} <$ 460 MeV;}
\item{the sum of lepton energies in the range:\\ 470 MeV $< E_{e^+} + \,E_{e^-} <$ 750 MeV;}
\item{the sum of the energies of the photons from $\pi^0$ decay in:\\ 300 MeV $< E_{\gamma_1} + E_{\gamma_2} <$ 670 MeV;}
\item{ all particles in the final state in the angular acceptance  $35^{\circ} < \theta < 135^{\circ}$;}
\item{the opening-angle between tracks and between the two photons being respectively $\theta_{e^+e^-}^{open} < 145^{\circ}$ and  $27^{\circ} < \theta_{\gamma_1\gamma_2}^{open} < 57^{\circ}$.}
\end{itemize}

The other relevant contribution to the background is from the radiative decay $\phi \to \pi^0 \gamma$, with the real final state photon converting to an $e^+e^-$ pair in the interaction with the beam-pipe (BP) and drift-chamber (DC) walls or with the $\pi^0$ going to a single Dalitz decay ($\pi^0 \to \gamma e^+e^- $). The contribution due to photon conversion can be suppressed cutting on the invariant-mass and distance between the tracks at BP and DC walls.

\begin{figure}[!h]
\begin{minipage}{0.51 \linewidth}
\centerline{\includegraphics[width=1.12\linewidth]{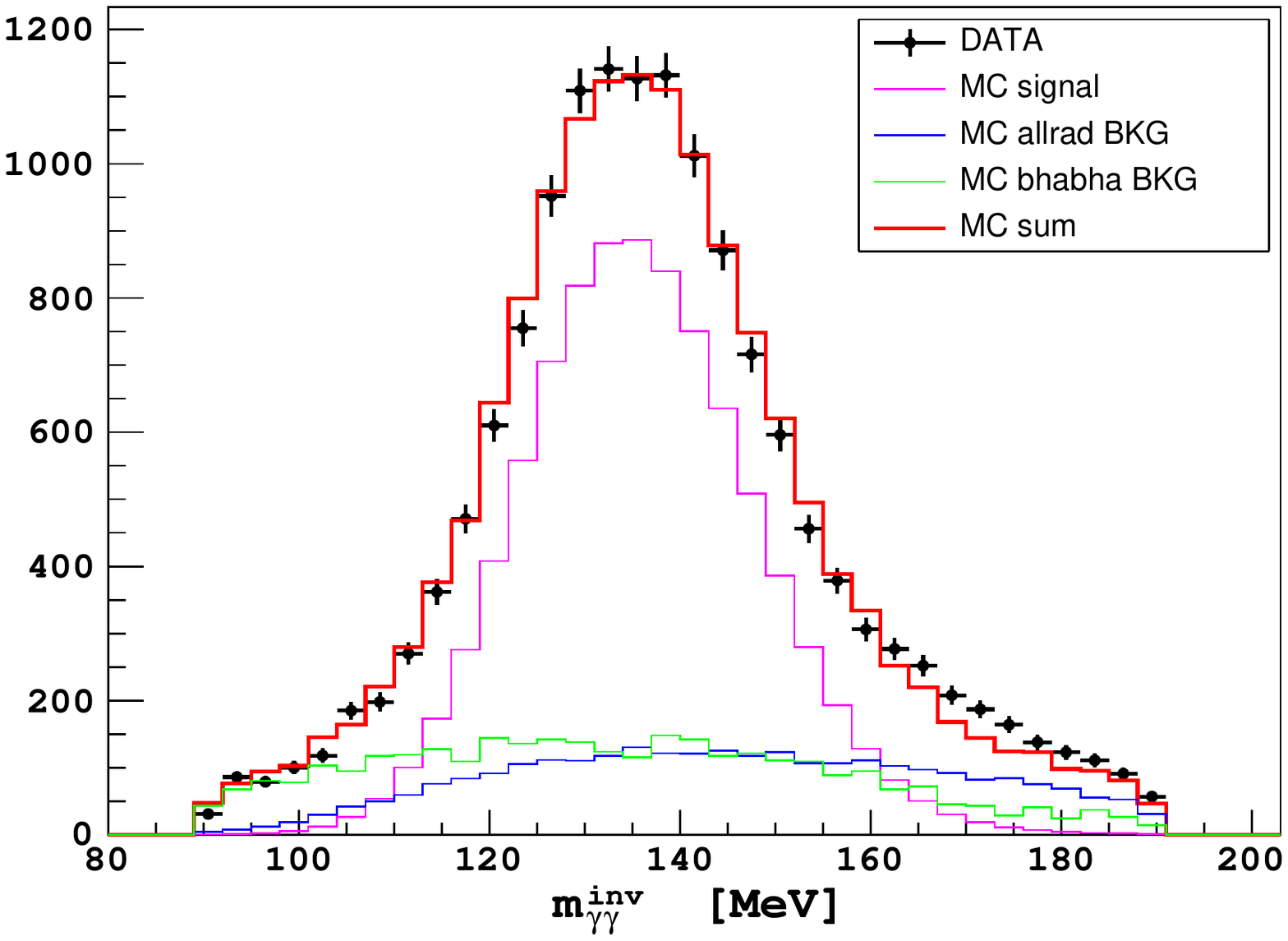}}
\end{minipage}
\begin{minipage}{0.51 \linewidth}
\centerline{\includegraphics[width=1.12\linewidth]{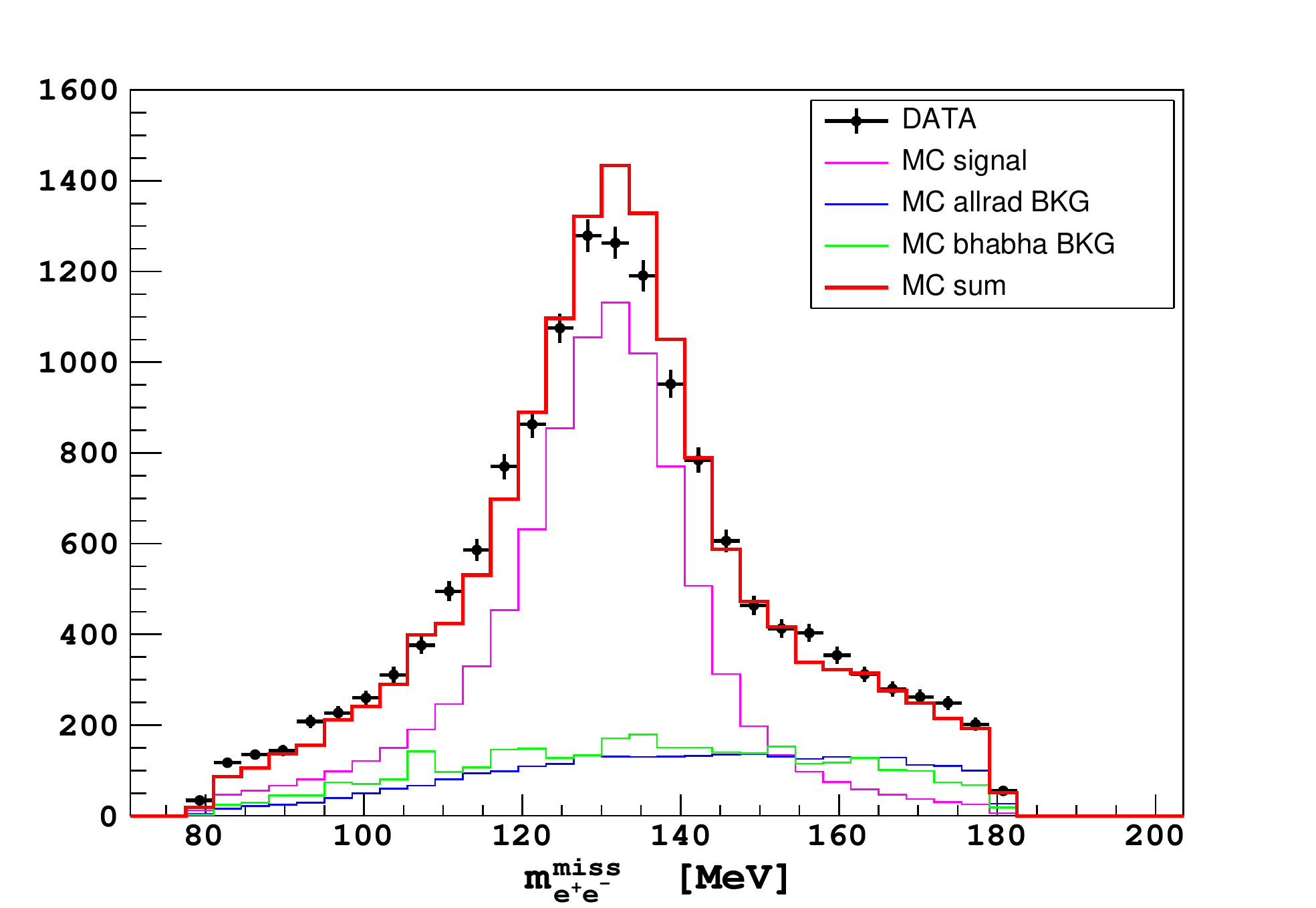}}
\end{minipage}
\begin{minipage}{0.51 \linewidth}
\centerline{\includegraphics[width=1.13\linewidth]{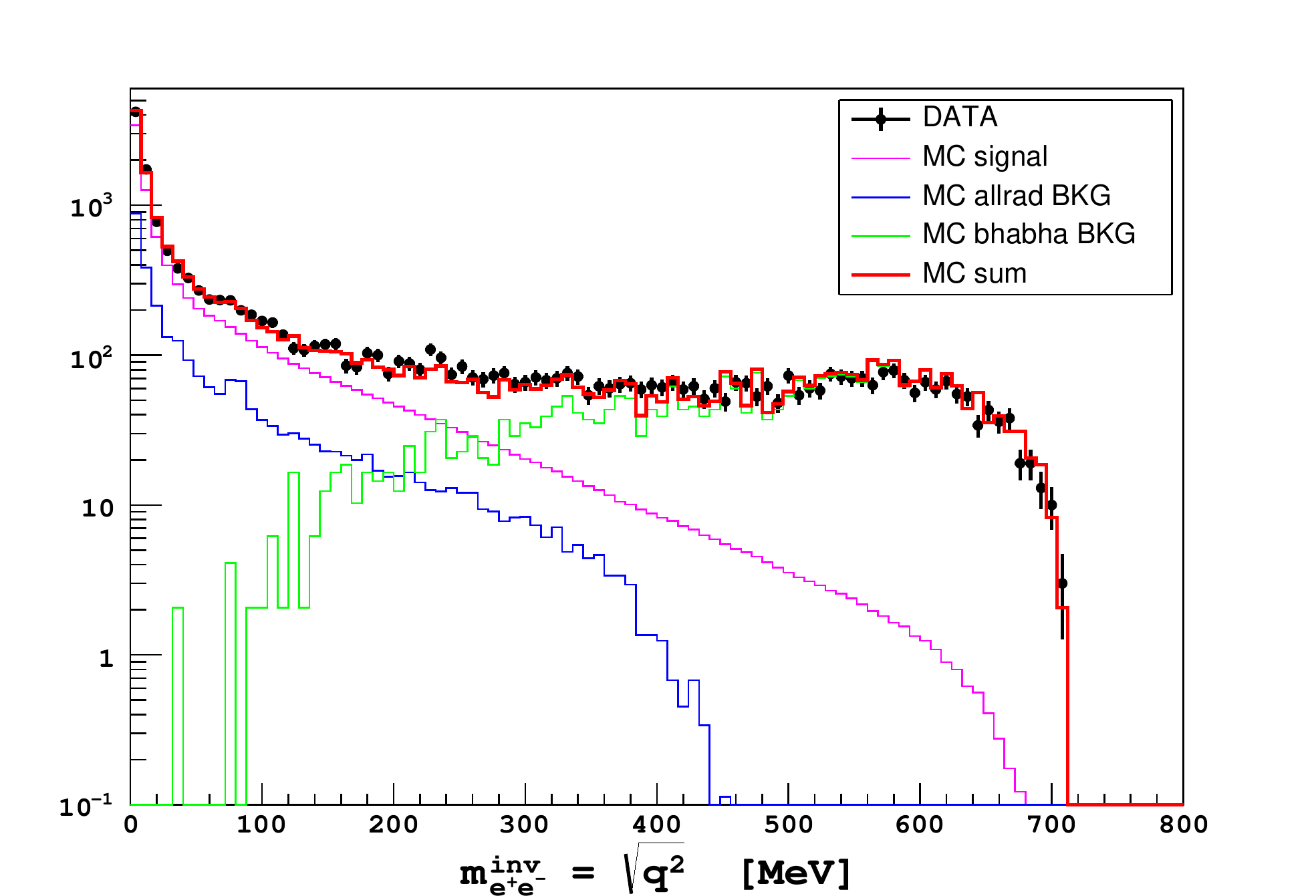}}
\end{minipage}
\begin{minipage}{0.51 \linewidth}
\centerline{\includegraphics[width=1.10\linewidth] {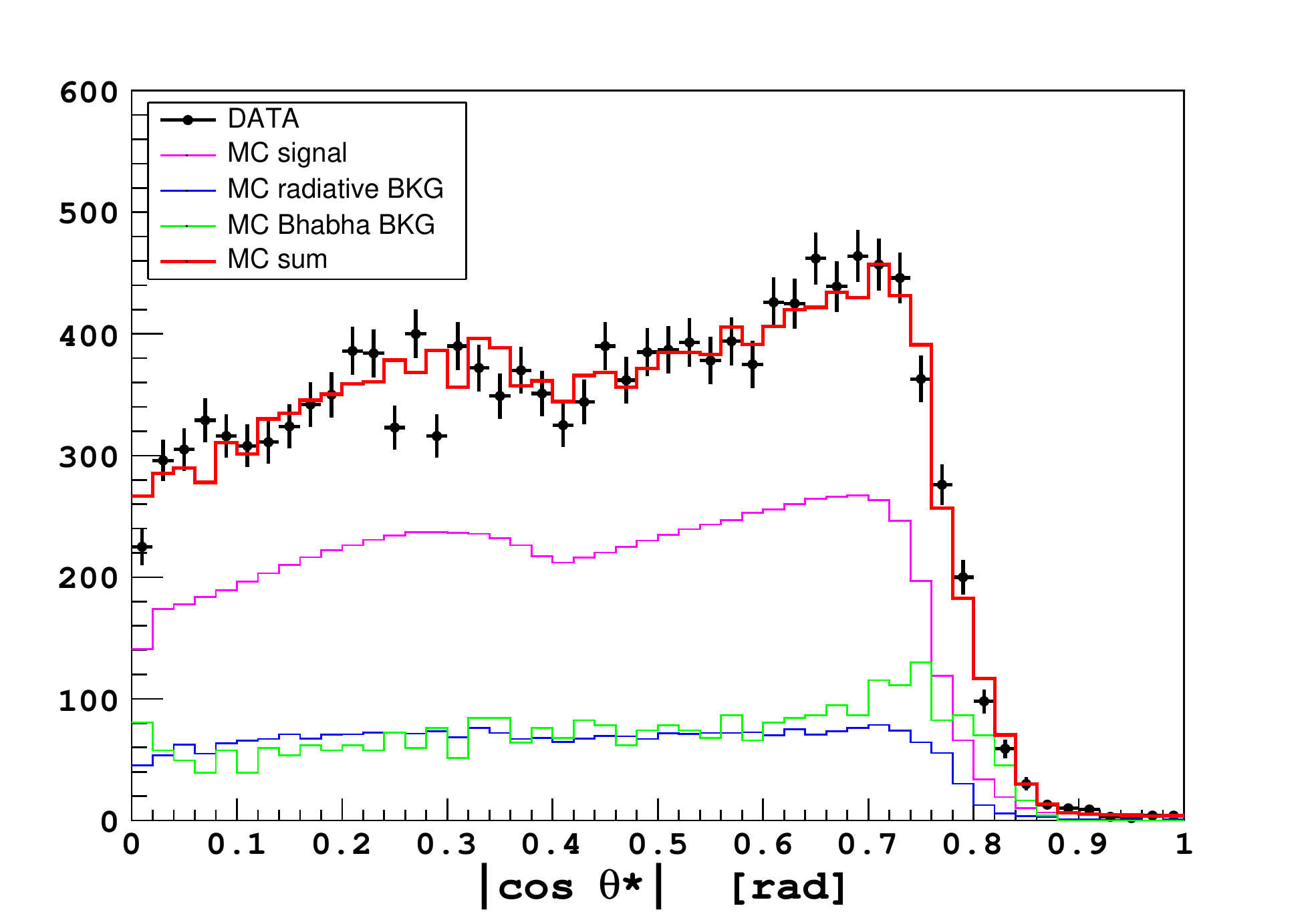}}
\end{minipage}
\caption[]{Comparison between data (black points) and MC distributions (red histogram is the MC sum) for: the invariant mass of the two photons (top left), the recoil mass against the $e^+e^-$ pair (top right), the $e^+e^-$ invariant-mass spectrum (bottom left) and the angle between the momentum direction of $\phi$ and the $e^+$ in the rest-frame of the lepton-pair (bottom right).}
\label{fig:dataMC}
\end{figure}

A further improvement of the signal to background ratio is achieved by asking for: the invariant-mass of the two prompt clusters to be within the window 90 MeV $< m_{\gamma\gamma}^{inv} <$ 190 MeV, the missing-mass to the tracks in the range: 80 MeV $< m_{e^+e^-}^{miss} <$ 180 MeV and the total invariant-mass of the four final-state particles to be compatible within 30 MeV with the $\phi$ meson mass.

The agreement between data and Monte Carlo simulation is shown in Fig.~\ref{fig:dataMC} for different kinematical variables.

At the end of the analysis, about 14500 events are selected, with a total background contamination of $\sim\!\!30 \%$. The background contribution is removed bin-by-bin\footnote{The bin width increases with increasing $\sqrt{q^2}$ in order to optimize the statistics of data and Monte Carlo samples.} by subtracting the fits to each single background component from data points in the $m_{e^+e^-}^{miss}$ distribution. To improve the fit procedure of the Monte Carlo shapes, a global fit to data is performed.  

\begin{figure}[htpb!] 
\centering
\includegraphics[width=0.95 \textwidth, trim= 0mm 0mm 0mm 0mm, clip]{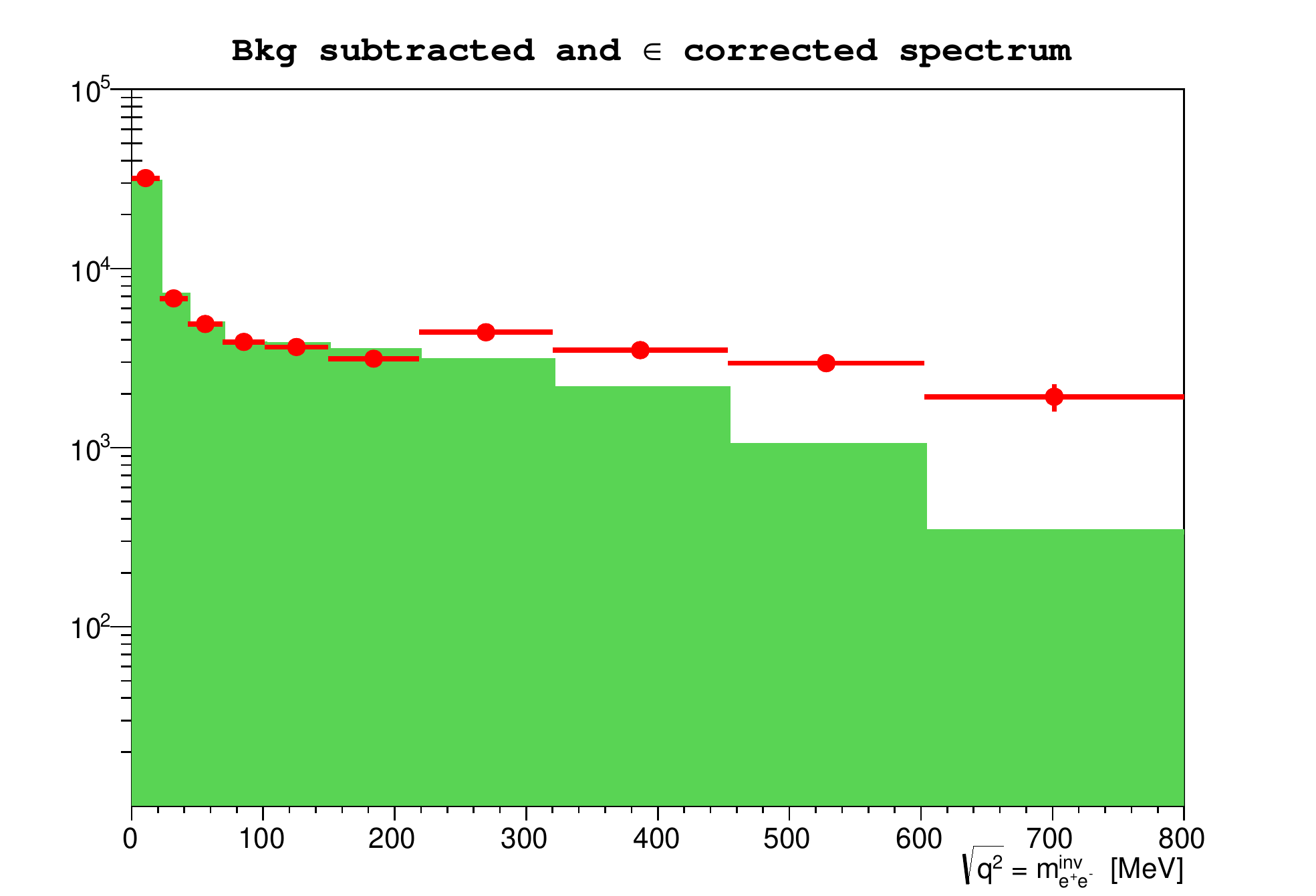}                
\caption[]{Preliminary background-subtracted and efficiency corrected $e^+e^-$ invariant-mass spectrum for 1.7 fb$^{-1}$ integrated luminosity (red dots). The green area corresponds to the expected MC distribution for a constant TFF.}
\label{fig:mee}                          	
\end{figure}

Fig.~\ref{fig:mee} shows a preliminary re-binned invariant-mass spectrum of $e^+e^-$, after the background subtraction and the efficiency correction (red points) as compared to the expected distribution for $|F_{\pi^0 \gamma^{*}}(q^2)|^2 = 1$. In the plot, only the statistical error is reported for data points. 

\section{Conclusions}

The status of the study of $\phi \to \pi^0 e^+e^-$ at KLOE was presented. The analysis, based on a data sample of 1.7 fb$^{-1}$ collected in 2004/2005 at $\sqrt{s} = m_{\phi}$, allowed the selection of about 9000 signal candidates, with a good agreement between data and MC in all kinematical variables. A deviation of data from the MC simulation (with a constant TFF parametrization) is observed at higher $\sqrt{q^2}$ of the $e^+e^-$ mass spectrum. This can be interpreted as the effect of a non-constant form factor playing a role in the decay.

Thanks to the statistics available for data, an improvement of a factor $\sim\!\!10$, with respect to the previous measurement of SND \cite{achasov} and CMD-2 \cite{akhm} experiments, is expected in the statistical error of the Branching Ratio measurement. 

The $F_{\phi \, \pi^0 \, \gamma^*}$ will be measured for the first time in this kinematical region; this will provide an strong consistency check of all theoretical model describing the TFF of the $\pi^0$ meson.  

\section{Acknowledgements}
We warmly thank our former KLOE colleagues for the access to the data collected during the KLOE data taking campaign.
We thank the DA$\Phi$NE team for their efforts in maintaining low background running conditions and their collaboration during all data taking. We want to thank our technical staff: 
G.F.~Fortugno and F.~Sborzacchi for their dedication in ensuring efficient operation of the KLOE computing facilities; 
M.~Anelli for his continuous attention to the gas system and detector safety; 
A.~Balla, M.~Gatta, G.~Corradi and G.~Papalino for electronics maintenance; 
M.~Santoni, G.~Paoluzzi and R.~Rosellini for general detector support; 
C.~Piscitelli for his help during major maintenance periods. 
This work was supported in part by the EU Integrated Infrastructure Initiative Hadron Physics Project under contract number RII3-CT- 2004-506078; by the European Commission under the 7$^{\mathrm{th}}$ Framework Programme through the `Research Infrastructures' action of the `Capacities' Programme, Call: FP7-INFRASTRUCTURES-2008-1, Grant Agreement No. 227431; by the Polish National Science Centre through the Grants No. 
DEC-2011/03/N/ST2/02641, 
2011/01/D/ST2/00748,
2011/03/N/ST2/02652,
2013/08/M/ST2/00323,
and by the Foundation for Polish Science through the MPD programme and the project HOMING PLUS BIS/2011-4/3.


\end{document}